\newcommand{\ACP}{A_{\rm CP}}

\newcommand{\EE}{e^+e^-}

\newcommand{\jpsi}{J/\psi}

\newcommand{\x}{X(3872)}

\newcommand{\LLb}{\Lambda\bar{\Lambda}}

\newcommand{\rt}{\rightarrow}

\documentclass[wssquare]{ws-procs961x669}  

\begin{document}
\title{Nucleon/Hyperon Physics at BES}

\author{Stephen Lars Olsen}

\address{University of Chinese Academy of Science,\\
Beijing 100039, CHINA\\
$^*$E-mail: solsensnu@gmail.com\\
}

\begin{abstract}
  The year 2019 marks the 30$^{\rm th}$ anniversary of BES and the 100$^{\rm th}$ anniversary
  of Rutherford's discovery of the proton.  In spite of the fact that when BES operations started
  the proton was already 70 years old and the strange hyperons were all over 25, BES continues
  to make important and unique measurements of nucleon and hyperon properties, including some interesting
  discoveries.
\end{abstract}


\bodymatter

\section{Introduction}\label{sec1:intro}
\begin{figure}[h]
\begin{center}
\includegraphics[width=4in]{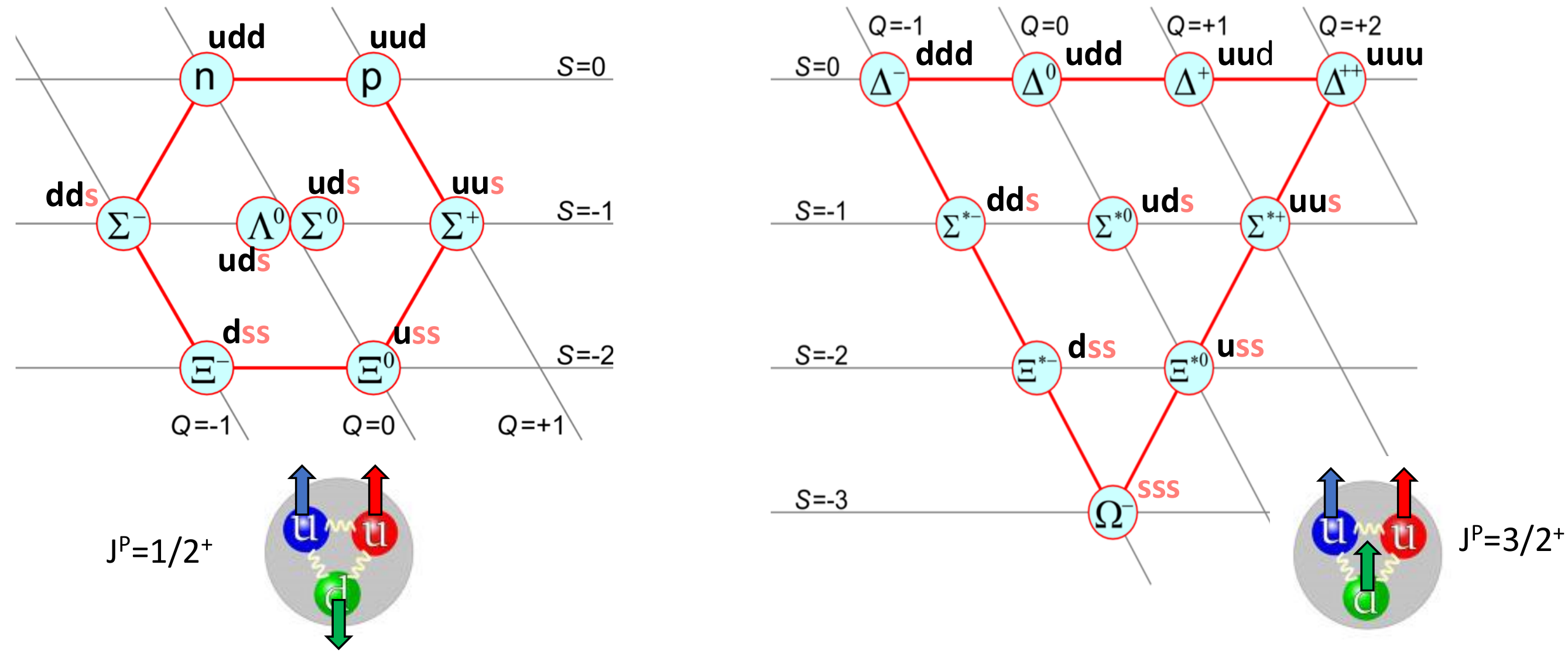}
\end{center}
\caption{The baryon octet (left) and decuplet (right). The horizontal axis is isospin, the vertical
  axis is strangeness, and the quark contents are indicated. All octet baryons have spin-parity $J^P=1/2^+$,
  nd the decuplet baryons have $J^P=3/^+$. Other than the stable proton and the $\Sigma^0$, which decays
  electromagnetically to $\gamma\Lambda$, the octet baryons decay via weak interactions. The $\Omega^-$
  is the only weakly decaying decuplet baryon.
  }
\label{fig:octet}
\end{figure}

\noindent
In 1919, Ernest Rutherford reported the observation of ``swift'' charged particles
produced by $\alpha$-particles impinging on $^{14}$N atoms (in modern parlence the
$^{14}$N($\alpha ,p$)$^{17}$O reaction). The swift particles
were positively charged and had a range in a zinc sulphide screen that
was ``far beyond the range of the $\alpha$ particles.'' He concluded that
these were hydrogen ions that had been ``constituent parts of the nitrogen nucleus,''
confirming the hypothesis of the chemist Willian Prout that the nuclei of all elements
contained hydrogen ions. The neutron, the isotopic spin partner of the proton, was
discovered 13 years later by James Chadwick in $\alpha$-particle beryllium collisions.
Since that time, the proton has been the prime subject of a huge amount of research,
including its anomalously large (compared to expectations for a Dirac point-like particle)
magnetic moment, $\mu_p$, first measured by Otto Stern in 1933, and its non-zero charge radius first
measured by Robert Hofstadter in 1956. The Lambda hyperon ($\Lambda$) was discovered via its production in
high energy cosmic rays in 1947. The $\Sigma^+,\ \Sigma^0,\ \Sigma^-$ and $\Xi^0,\ \Xi^-$ hyperons were
discovered at Brookhaven and Berkely in the 1950s, and the $\Omega^-$ at Brookhaven in 1964.
These particles and their properties provided essential clues for Gell-Mann's discovery of
flavor-$SU(3)$ and the Gell-Mann/Zweig quark model (see Fig.~\ref{fig:octet}). Most of these
particles decay via weak interactions, and measurements of their parity-violating weak decay
parameters were major experimental activities in the 1960s. By 1989, when the BES experimental
program started, this was already ancient history. Nevertheless BES, an experiment specialized for
$\tau$-charm physics, has managed to contribute substantially to our understanding of the well
established nucleons and strange hyperons.

\section{Space-like baryon form-factors}
\noindent
Most of what we know about the internal structure of baryons comes from a huge number of experiments that
measured the elastic scattering of high energy electrons from proton and, to a lesser extent, neutron targets.
For single photon exchange, the laboratory differential cross section is given by

\begin{equation}
  \frac{d\sigma}{d\Omega} = \frac{\alpha^2}{4E^2_e \sin^4\frac{\theta}{2}}\frac{E_e^{\prime}}{E_e}
  \Big(\frac{G^2_E(Q^2) +\tau G^2_M(Q^2)}{1+\tau}\cos^2\frac{\theta}{2}+2\tau G^2_M(Q^2)\sin^2\frac{\theta}{2}\Big),
  \label{eqn:space-like}
\end{equation}
\noindent
where $E_e$ ($E_e^{\prime}$) is the incident (scattered) electron energy, $\theta$ is the scattering
angle, $Q^2=-q^2=(p_e-p_e^{\prime})^2$ and $\tau=Q^2/4m^2_p$; $q^2$ is the exchanged photon's squared invariant
mass, which in scattering experiments is always {\it space-like} (i.e., $q^2<0 $).

\begin{figure}[h]
\begin{center}
\includegraphics[width=4.0in]{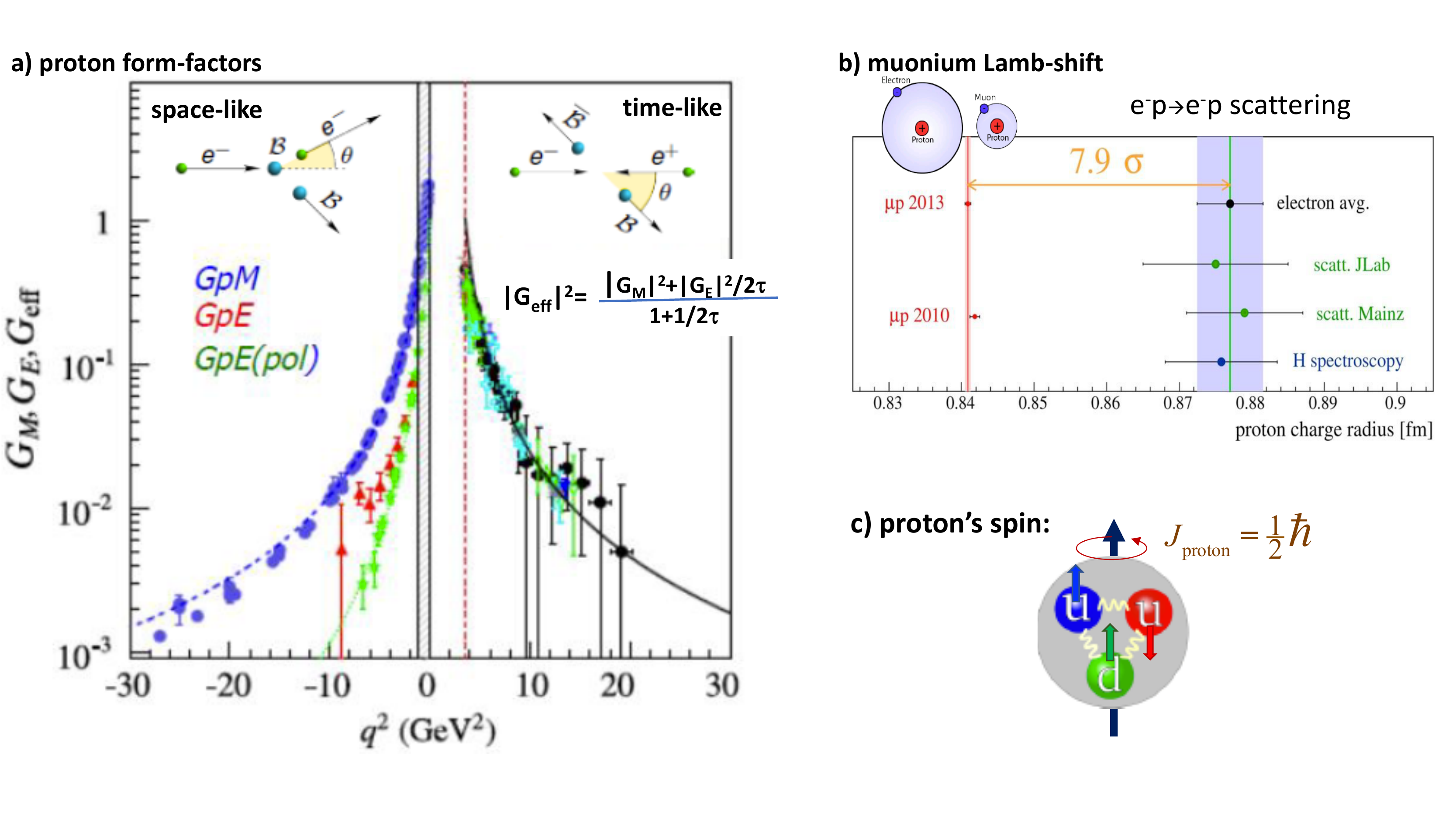}
\end{center}
\caption{{\bf a)} Proton form-factor measurements (from ref.~\cite{pacetti}). {\bf b)} Comparison of
  $r^{\rm rms}_{ff}$ (blue band) and $r^{\rm rms}_{\mu^- p}$ (red stripe) measurements. {\bf c)} The proton's
  spin in the na\"{i}ve quark model.
 }
\label{fig:form-factors}
\end{figure}

The functions $G_E(Q^2)$ and $G_M(Q^2)$ are the baryon's electric and magnetic form-factors that characterize
the proton's charge and current densities. These are real, analytic functions: at $Q^2 \rt 0$, $G_E \rt 1$ and
$G_M \rt \mu_p$.  The proton's electric and magnetic form-factors have been measured in hundreds of experiments;
averaged $G_E(Q^2)$ and $|G_M(Q^2)|$ measurements from recent experiments are shown on the space-like side of
Fig.~\ref{fig:form-factors}a.
These measurements were used to infer the  proton's rms charge-radius with impressive precision:
$r^{\rm rms}_{ff}=0.8775\pm 0.0005$~fm.  However, this value disagrees rather violently with
recent charge-radius determinations based on the Lamb shift of muonium (i.e., $\mu^-$p atoms):
$r^{\rm rms}_{\mu^- p}=0.842\pm 0.001$~fm (Fig.~\ref{fig:form-factors}b). Thus, in spite of 60 years
of measurements, we still do not understand the way charge and currents are configured in the proton.

In its most na\"{i}ve form, the quark model has the proton comprised of three spin-1/2 quarks in $S$-waves,
where two of them have $S=0$ with the third one providing the proton's $J=\hbar/2$ net angular momentum
(Fig.~\ref{fig:form-factors}c). This can be tested with asymmetry measurements for polarized electron beams
scattering from polarized proton targets. The green points on Fig.~\ref{fig:form-factors}a's space-like
side show the resulting polarization form-factor $|G_{E(pol)}|$ values determined from such measurements.
In somewhat of a surprize, these polarization measurements show that the net quark-spin contribution to the
proton's total spin is about $0.15\hbar$, or only a third of the proton's total angular momentum; the rest
must come from gluons and/or orbital angular momentum. This was totally unanticipated and is further evidence
of the paucity of our understanding of the proton.

\section{Time-like baryon form-factors}
When faced with a challenging puzzle, what can an experimenter do? ...the BESIII answer is do more
experiments. While electron-positron colliders cannot access space-like baryon form-factors, they are extremely
well suited to measure the analytic continuations of $G_E$ and $G_M$ into the time-like region ($Q^2=q^2>0$) via
$\EE\rt B\bar{B}$ interactions. This is especially true for BESIII, since the BEPCII CM energy range includes
the thresholds for the pair production of all of the stable baryons. The time-like equivalent of
eqn.~\ref{eqn:space-like} is
\begin{equation}
  \frac{d\sigma}{d\Omega} = \frac{\alpha^2\beta {\mathcal C}}{4E^2_{CM}}
  \Big((1+\cos^2\Theta)|G_M(Q^2)|^2 +\frac{\sin^2\Theta}{\tau}|G^2_E(Q^2)|^2\Big),~~~\beta=\sqrt{1-\frac{1}{\tau}},
  \label{eqn:time-like}
\end{equation}
\noindent
where $\beta$ and $\Theta$ are the baryon's CM velocity and production angle, and
${\mathcal C}$ accounts for the effects of Coulomb interactions between the baryons: for
point-like charged particles, ${\mathcal C}=\pi\alpha/[\beta(1-\exp(-\pi\alpha/\beta)]$; for
point-like neutral particles ${\mathcal C}=1$.

It is evident from  eqn.~\ref{eqn:time-like} that separate measurements of $|G_E|$ and $|G_M|$ can be extracted
from the $\cos\Theta$ dependence of the differential cross section. However, in contrast to space-like form-factor
measurements, existing data on time-like form-factors are very limited and, prior to recent BaBar and BESIII
measurements, separate determinations of $G_E$ and $G_M$ of any significance had not been possible. Instead
measurements have focused on the magnitude of the ``effective'' form-factor
$|G_{\rm eff}|^2 = (|G_E|^2+2\tau|G_M|^2)/(2\tau +1)$, and this is what is shown on the time-like side of
Fig.~\ref{fig:form-factors}a. Analyticity requires that at threshold $R\equiv|G_E(4m^2_B)/G_E(4m^2_B)|=1$, and
separate $|G_E|$ and $|G_M|$ measurements near threshold could be used to test this relation. Unlike space-like
form-factors, which both have to be real, time-like form-factors can be complex. A non-zero relative complex
phase between $G_E$ and $G_M$ ($\Delta\Phi$) would result in polarization of
the final-state baryons. Measurements of this polarization for $\EE\rt p\bar{p}$ are impractical with BESIII.
However, the large parity-violating weak decay asymmetry parameters for the $\Lambda$, $\Sigma^+$ and
$\Xi$ hyperons make BESIII uniquely well suited for $\Delta \Phi$ measurements for these baryons.

\subsection{Near-threshold $\EE\rt B\bar{B}$ cross section measurements}
\noindent

For point-like charged particles at and just above threshold ($E_{CM} = 2m_B$),
${\mathcal C}\approx \pi\alpha/\beta$; the $\beta$ in the denominator of ${\mathcal C}$ cancels the
$\beta$ in the numerator of eqn.~\ref{eqn:time-like} and the cross section is $\pi^2\alpha^3/2m^2_B $, and
non-zero right at threshold. Thus, for example, the cross section for $\EE\rt\tau^+\tau^-$ jumps from zero
to $\delta\sigma(m_\tau)=\pi^2\alpha^3/2m^2_\tau = 237$~pb right at threshold. (This cross-section relation,
with its 237~pb threshold jump, is used for all of the BES $\tau$-mass measurements.) This abrupt jump
reflects the influence of the infinite number of Bohr-like Coulomb bound $\tau^+\tau^-$ states that
are just below the $2m_{\tau}$ threshold. For neutral point-like particles, ${\mathcal C}=1$ and there is no
cancelation, and the cross section starts at zero and increases as $\sigma\propto\beta$.

\begin{figure}[h]
\begin{center}
\includegraphics[width=3.8in]{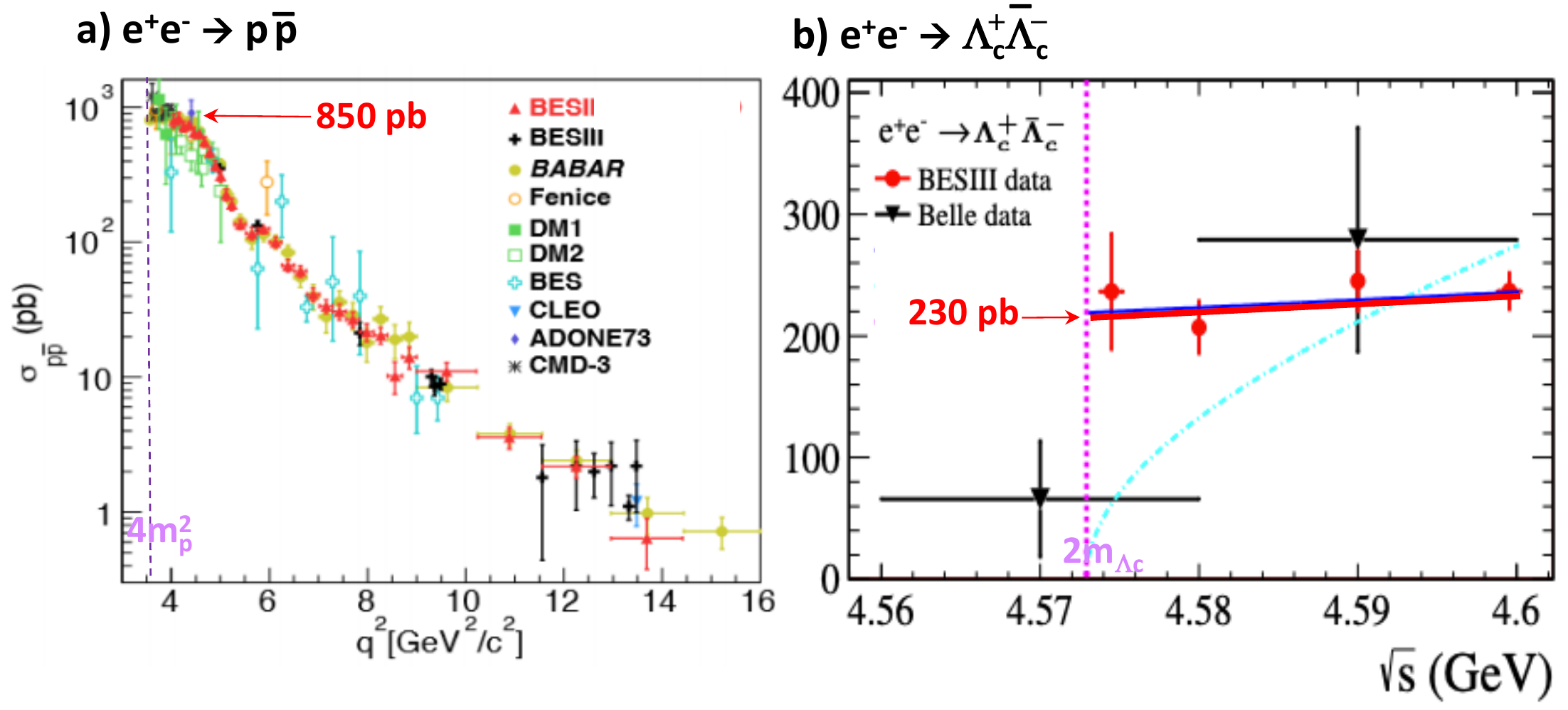}
\end{center}
\caption{Near-threshold behavior of  {\bf a)} $\sigma(\EE\rt p\bar{p})$ and
  {\bf b)} $\sigma(\EE\rt\Lambda^+_c \bar{\Lambda}_c^-)$.
 }
\label{fig:sigma-ppbar-LcLcbar}
\end{figure}

In the $\EE\rt p\bar{p}$ reaction at threshold, the $p\bar{p}$ pair is produced via a $Q^2\approx 3.5$~GeV$^2$
virtual photon, with a Compton wavelength that is $\approx 0.1$~fm -- one-tenth the proton's rms size. On this
distance scale, the proton is definitely not point-like and the expression in eqn.~\ref{eqn:time-like} should
not apply. Thus it was a big surprize that near-threshold $\sigma(\EE\rt p\bar{p})$ measurements from
BaBar~\cite{BaBar-ppbar}, subsequently confirmed by CMD3~\cite{CMD3-ppbar} and
BESIII~\cite{BESIII-ppbar,BESIII-ppbar-isr}, show a dramatic jump at threshold that is consistent with the
$\delta\sigma(m_p)=\pi^2\alpha^3/2m^2_p = 850$~pb value that would be expected for a point-like proton
(see Fig.~\ref{fig:sigma-ppbar-LcLcbar}a).  The CMD3 measurements are particularly notable in that they
span the $Q^2=4m^2_p$ threshold in small $E_{\rm CM}$ increments and find that the ``jump'' occurs across less than
a 1~MeV change in CM energy. BESIII recently reported a similar behavior for
$\sigma(\EE\rt \Lambda^+_c \bar{\Lambda}^-_c)$ near the $E_{\rm CM}=2m_{\Lambda_c}$ threshold~\cite{BESIII-LcLcbar}, shown
in Fig.~\ref{fig:sigma-ppbar-LcLcbar}b, where a measured $230\pm 50$~pb threshold cross section jump is even
larger than expectations for a charged point-like particle, i.e., $\delta\sigma(m_{\Lambda_c})=140$~pb.

\begin{figure}[h]
\begin{center}
\includegraphics[width=3.8in]{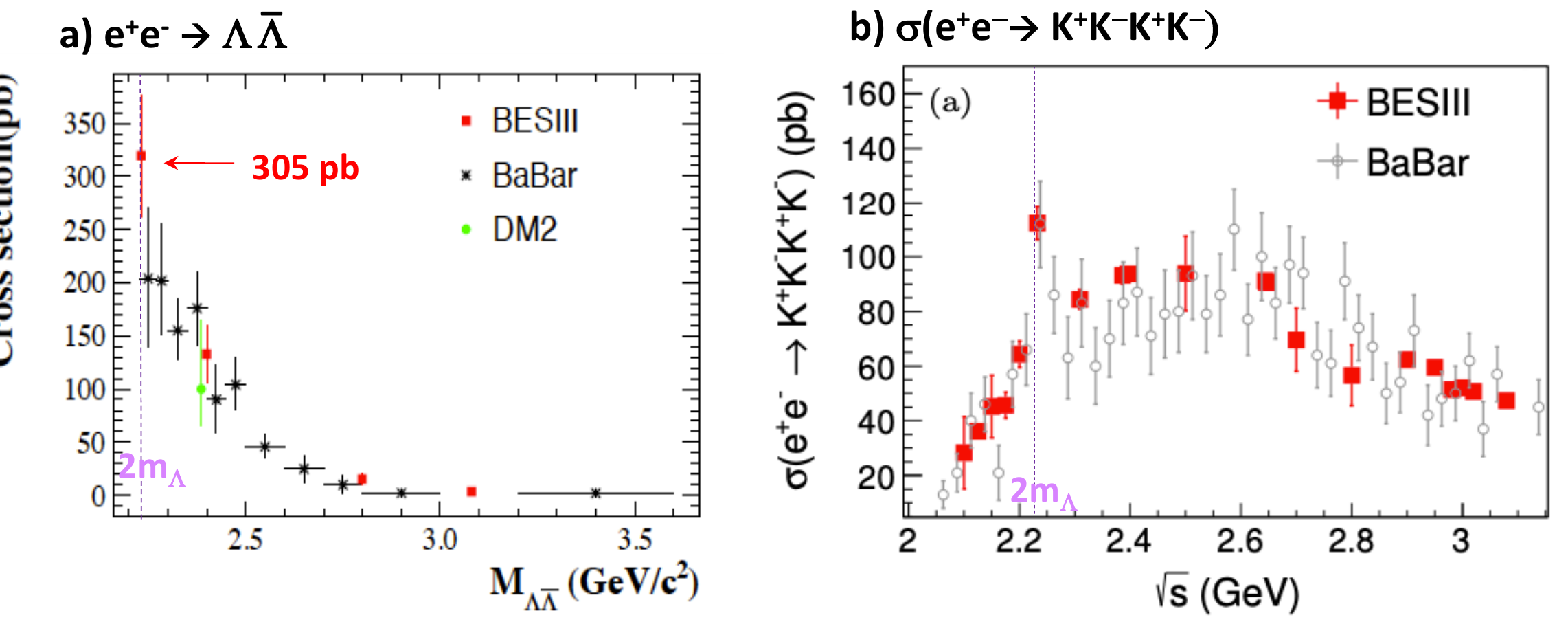}
\end{center}
\caption{Near-threshold behavior of 
  {\bf a)} $\sigma(\EE\rt\Lambda \bar{\Lambda})$ and
  {\bf b)} BESIII and BaBar measurements of $\sigma(\EE\rt K^+K^-K^+K^-)$.
}
\label{fig:sigma-LLbar-SpSmbar}
\end{figure}

For neutral point-like particles, the threshold cross-section is expected to be zero and the
increase as $\sigma\propto\beta_\Lambda$.  However, while measured $p\bar{p}$ and $\Lambda_c^+\bar{\Lambda}_c^-$
cross-sections seem to reflect point-like charged-particle behavior, BESIII results on
$\sigma( \EE\rt\Lambda\bar{\Lambda})$, see Fig.~\ref{fig:sigma-LLbar-SpSmbar}a, do not follow
expectations for a neutral point-like particle. These measurements, which start at $E_{\rm CM}=2m_\Lambda+1.0$~MeV
($\beta\approx 0.03$), find a non-zero threshold value of $305\pm 50$~pb that decreases with increasing
CM energy~\cite{BESIII-LLbar}. This jump is about half of the $\delta\sigma(m_\Lambda)=600$~pb expectation for
a point-like charged particle with the $\Lambda$ mass.

\subsubsection{Influence of sub-threshold baryonium states?}
\noindent
Theoretical predictions for $\sigma(\EE\rt p\bar{p})$~\cite{Meissner-ppbar} and
$\sigma(\EE\rt \Lambda\bar{\Lambda})$~\cite{Meissner-LLbar} based on chiral effective theory fail to
reproduce the very sharp threshold jumps, especially those with less than a 1~MeV turn-on as seen for
$p\bar{p}$ by CMD3 and $\Lambda\bar{\Lambda}$ by BESIII.  For the $\Lambda\bar{\Lambda}$ case, the authors of
ref.~\cite{Meissner-LLbar} suggested that this anomalous threshold behavior could be the influence of a
narrow $^3S_1$ $\Lambda\bar{\Lambda}$ bound state with mass very near the $2m_\Lambda$ threshold. Intriguingly,
measurements of $\sigma(\EE\rt K^+K^-K^+K^-)$ by both BESIII and BaBar, shown in
Fig.~\ref{fig:sigma-LLbar-SpSmbar}c, have, respectively, $\sim 3\sigma$
and $\sim 2\sigma$ enhancements in the CM energy  bin that includes $E_{\rm CM}= 2m_{\Lambda}$.

Perhaps the observed threshold behavior in other channels are also due to near-threshold $^3S_1$ $B\bar{B}$
bound states.  In 2003, BESIII reported the discovery of a strong candidate for a
sub-threshold $^1S_0$ $p\bar{p}$ bound state in $\jpsi\rt \gamma p\bar{p}$ decays~\cite{BESII-X1860}, as
discussed in detail at this symposium by Prof.~Shan Jin.
Thus, it does not seem unreasonable to think that a corresponding $^3S_1$ state might exist.
These results suggest that extended BESIII operations in the vicinities of all of the stable baryon-antibaryon
thresholds might yield very interesting results. 

\subsection{Time-like form-factor measurements at higher $Q^2$ values}
\noindent
In addition to the near-threshold measurements described in the previous subsection, BESIII has performed
above-threshold measurements of $\EE\rt p\bar{p}$ and $\EE\rt\Lambda\bar{\Lambda}$.  The former have yielded
precise separate measurements of $|G_E|$ and $|G_M|$ for the proton; the latter have resulted in
the first-ever measurement of a non-zero $\Delta\Phi$ value for any baryon.

\subsubsection{$\EE\rt p\bar{p}$}
\begin{figure}[h]
\begin{center}
\includegraphics[width=5.0in]{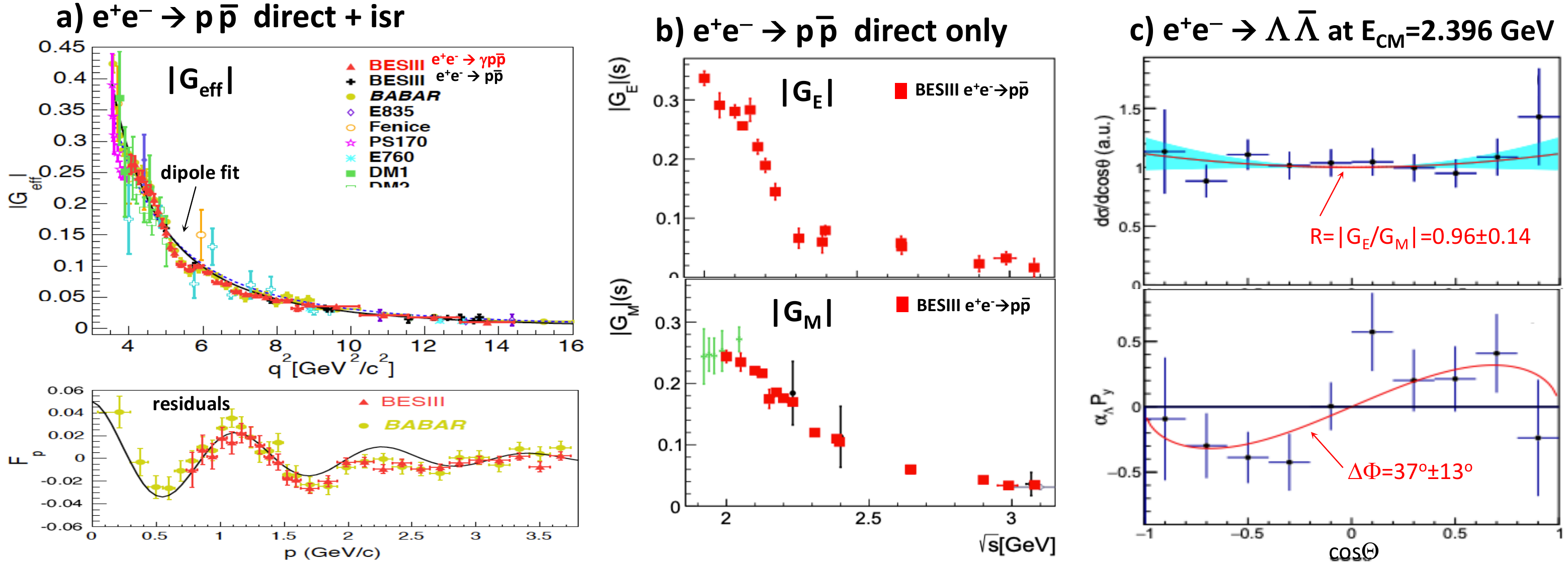}
\end{center}
\caption{
  {\bf a)}  The top panel shows measurements of $|G_{\rm eff}(Q^2)|$ of the proton. The
  black curve is the result of a fit with a simple dipole form. The bottom panel shows the residuals from
  the dipole fit. {\bf b)} The red points show BESIII $|G_E(Q^2)|$ (top) and $|G_M(Q^2)|$ (bottom) measurements.
  {\bf c)} BESIII $d\sigma /d\cos\Theta$ measurements for $\EE\rt\Lambda\bar{\Lambda}$ at $E_{\rm CM}=2.396$~GeV
  (top) and the $\cos\Theta$ dependence of the $\Lambda$ polarization (bottom). The red curves are fits to the
  data.
}
\label{fig:ppbarff-Lambda-Pol}
\end{figure}

BESIII measured the proton's time-like proton form-factor using $\EE\rt p\bar{p}$ events collected at
12~CM~energy points ranging from 2.232~GeV to 3.671~GeV~\cite{BESIII-ppbar} and $\EE\rt \gamma p\bar{p}$
initial-state-radiation (isr) events in large data sets with $E_{\rm CM}$ between
3.773~and~4.600~GeV~\cite{BESIII-ppbar-isr}. In the isr events, the $\gamma$-rays are preferentially emitted at
small angles and do not register in the detector; their presence and four-momenta are inferred from
energy-momentum conservation. Figure~\ref{fig:ppbarff-Lambda-Pol}a (top) shows BESIII $|G_{\rm eff}|$
measurements along with those from previous experiments.  The black curve shows the result of
a fit to a dipole shape ($|G_{\rm eff}|\propto 1/[(1+Q^2/m^2_a)(1-Q^2/Q^2_0)]$) that is commonly used to
characterize form-factors. Oscillatory deviations from the dipole shape, first noted by
BaBar~\cite{BaBar-ppbar}, are pronounced, as indicated by the fit residuals shown in the lower panel of
Fig.~\ref{fig:ppbarff-Lambda-Pol}a. These deviations may be the indication of broad $J^{PC}=1^{- -}$ resonances
in the $p\bar{p}$ channel or some unanticipated dynamics in baryon-antibaryon systems~\cite{ppbar-oscill}.
In either case, they require further study both in the nucleon-antinucleon and hyperon-antihyperon channels.

The isr measurements have the advantage of providing access to a broad range of $Q^2$ at once, and can
be done with the large samples of BESIII data that are taken at $E_{\rm CM}=3.77$~GeV for $D$-meson decay
measurements and $E_{\rm CM}>4.0$~GeV for $XYZ$-meson studies. However, the required emission of a hard isr
$\gamma$-ray incurs a luminosity penalty of order $\alpha_{\rm QCD}/\pi$.  Thus, for BESIII at least, detailed
form-factor measurements, including precise separate determinations of $|G_E|$ \& $|G_M|$, and measurements
of $\Delta \Phi$, their relative phase, are best done with direct $\EE\rt B\bar{B}$ production events and
this reqires machine-time-consuming dedicated runs.

The BESIII direct $\EE\rt p\bar{p}$ measurements~\cite{BESIII-ppbar} were used to make the precison
extractions of $|G_E$ and $|G_M|$ shown, respectively, in the top and bottom panels of
Fig.~\ref{fig:ppbarff-Lambda-Pol}b. Here, for the point nearest threshold (at $E_{\rm CM}=2.0$~GeV),
$|G_E|\neq|G_M|$; BESIII measures $R(Q^2=4.0\ {\rm GeV}^2)=1.38\pm 0.11$, in agreement with a
previous BaBar measurement ($1.48\pm 0.16$)~\cite{BaBar-ppbar}, with better precision. Analyticity
requires $R=1$ at $Q^2=4m_p^2=3.52$~GeV$^2$, and this may be in trouble 
unless there are rather dramatic changes in the $|G_E|$ and $|G_M|$ trends for $Q^2<4.0$~GeV$^2$.

With data collected at $E_{\rm CM}=2.396$~GeV, 165~MeV above the $\Lambda\bar{\Lambda}$ threshold, BESIII
used a low background sample of 555~$\EE\rt\Lambda\bar{\Lambda}$ events to make first
``complete'' time-like form-factor measurements for any baryon: $G_{\rm eff}=0.123\pm 0.004$; the
$d\Gamma/d\cos\Theta$, distribution, where $\Theta$ is the $\Lambda$ production angle,
is shown in Fig.~\ref{fig:ppbarff-Lambda-Pol}c (top), is consistent with being flat and
$R=|G_E/G_M|=0.96\pm 0.14$; the bottom panel of the figure shows strong evidence for a non-zero
$\cos\Theta$-dependent polarization corresponding to a  $|G_E|$-$|G_M|$ phase difference
of $\Delta\Phi=37^o\pm 11^o$. 

\section{Search for $CP$ violation in hyperon decays}

The CKM mechanism for $CP$ violation in the Standard Model (SM) fails to explain the matter-antimatter
asymmetry of the universe by more than 10 orders-of-magnitude. This suggests that additional,
heretofore undiscovered, $CP$ violating processes occur, and motivates aggressive searches for
new, non-SM sources of $CPV$. The LHCb experiment recently reported the first observation of
$CPV$ in charmed $D$-meson decays at the $10^{-3}$ level~\cite{Aaij}, which lies in the upper
range of SM expectations. To date, $CP$ violation in hyperon decays have never been observed. Standard Model
$CP$ violations in hyperon decays are expected to be $\sim 10^{-5}$; any value higher than this level
would be a signature of new, beyond the SM physics~\cite{pakvasa}. 

\subsection{BESIII limit on hyperon $CPV$ using $\jpsi\rt \Lambda\bar{\Lambda}$ events}
\noindent
The branching fraction for  $\jpsi\rt \Lambda\bar{\Lambda}$ is $1.94\pm 0.3 \times 10^{-3}$, which
translates into about $3.2\times 10^{6}$ fully reconstructed events with $\Lambda\rt p\pi^-$ and
$\bar{\Lambda}\rt \bar{p}\pi^+$ in BESIII's existing 10B $\jpsi$ event data sample.  These events are
jewels: they are essentially background-free and the $\Lambda$ and $\bar{\Lambda}$ are quantum-entangled
with completely correlated spin orientations. In the limit of strict $CP$ symmetry, the decay asymmetry
parameters for $\Lambda\rt p\pi^-$ $(\alpha_-)$ and  $\bar{\Lambda}\rt \bar{p}\pi^+$ $(\alpha_+ )$ are
equal in magnitude but opposite in sign, i.e. $\alpha_- = -\alpha_+$; deviations from this equality would be
a clear sign of $CP$ violation in hyperon decays. 

In $\jpsi\rt\Lambda\bar{\Lambda}$ decay, the $\Lambda\bar{\Lambda}$ pair is produced with either
parallel or antiparallel helicities. Prior to the BESIII discovery of non-zero $\Lambda$ polarization in
the form-factor measurement discussed in the previous section~\cite{Lambda-ff}, it was thought
that for $\jpsi$ decay, the amplitudes for the two helicity configurations were both real and the $\Lambda$
and $\bar{\Lambda}$ unpolarized.  In this case, only the product $\alpha_- \alpha_+$ would be
measureable, precluding the possibilty of a test of $CP$ symmetry. However the ref.~\cite{Lambda-ff}
results motivated BESIII to reexamine the case for $\Lambda$ polarization in $\jpsi$ decays~\cite{faldt}. 

\begin{figure}[h]
\begin{center}
\includegraphics[width=4.7in]{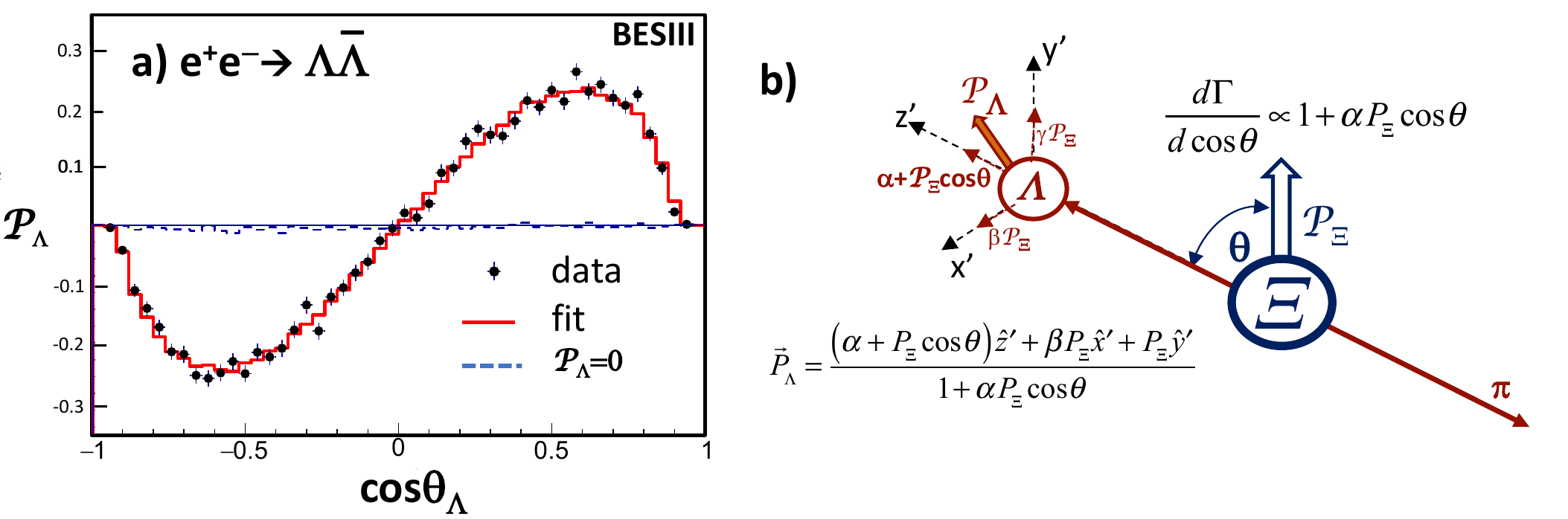}
\end{center}
\caption{
  {\bf a)} The $\cos\Theta_\Lambda$-dependent $\Lambda$ polarization in $\EE\rt \jpsi\rt \Lambda\bar{\Lambda}$,
  where $\Theta_\Lambda$ is the $\Lambda$ production angle.
  The red histogram is a fit to the data.
  {\bf b)} The $\Lambda$ angular distribution and polarization direction relative to the $\Xi$
  polarization direction for $\Xi\rt\Lambda\pi$ decay. Here $\alpha, \beta, \gamma$ refer to the $\Xi$
  weak decay parameters and $\theta$ is the angle between the $\Lambda$ momentum and $\Xi$ polarization.
}
\label{fig:Xi_2_Lambda}
\end{figure}

With 1.3B $\jpsi$ events collected in 2009 and 2012, a total of
420K fully reconstructed $\jpsi\to \LLb$ events with $\Lambda\to p\pi^-$
and $\bar{\Lambda}\to \bar{p}\pi^+$ were found. Fits to the data require 
a large relative phase between the two helicity amplitudes:
$\Delta\Phi=42.4^o\pm 0.8^o$~\cite{Lambda-CPV}, and a $\cos\Theta$-dependent
$\Lambda$ ($\bar{\Lambda}$) transverse polarization, shown in Fig.~\ref{fig:Xi_2_Lambda}a, that
varies between $-$25\% and $+$25\%.  Although the $\Lambda$ polarization averaged over $\cos\Theta$
is zero, the BESIII analysis uses the event-by-event polarization,
which has the average value $\langle |\vec{\mathcal P}_\Lambda| \rangle \approx 13\%$. This polarization
enables independent determinations of $\alpha_-$ and $\alpha_+$ and a test of the $CP$ relation.

For $\Lambda\to p\pi^-$, BESIII measured $\alpha_-=0.750\pm 0.010$, which is
more than five standard deviations higher than the previous world average
value of $\alpha_-=0.642\pm 0.013$~\cite{pdg2016} that was based entirely on pre-1974
measurements. The measured $\bar{\Lambda}\to \bar{p}\pi^+$  asymmetry parameter, 
$\alpha_+=-0.758\pm 0.012$, is also high, and consistent, within uncertainties, with
the $\alpha_-$ result. The $CP$ asymmetry,
$\ACP\equiv {(\alpha_-+\alpha_+)}/{(\alpha_--\alpha_+) } =-0.006\pm 0.014$, is
compatible with zero and a factor of two more sensitive than the best previous
$\Lambda$-based measurement, $\ACP=+0.013 \pm 0.022$~\cite{barnes}.  These results
were based on only $\sim 1/8^{\rm th}$ of BESIII's current 10B $\jpsi$ event sample.
Extending the ref.~\cite{Lambda-CPV} analysis to the full data sample should
improve BESIII's $CPV$ sensitivity by at least a factor of three.

\subsection{Prospects for hyperon $CPV$ searches using $\jpsi\rt\Xi\bar{\Xi}$ decays}
\noindent
BESIII's $CPV$ sensitivity will be further improved, potentially rather substantially, by using
polarized $\Lambda$ daughters from $\Xi\rt\Lambda\pi$ decays, where the $\Xi$s are produced via
$\jpsi\rt\Xi\bar{\Xi}$ ($\Xi=\Xi^-\ {\rm or}\ \Xi^0$). Figure~\ref{fig:Xi_2_Lambda}b shows the
$\Lambda$ polarization $\vec{\mathcal P}$ in $\Xi\rt\Lambda\pi$ decay. Of particular note is its
component along $\hat{z}^\prime$, the $\Lambda$ flight direction:
${\mathcal P}_{z^\prime}=\alpha + {\mathcal P}_\Xi\cos\theta$.  Here $ {\mathcal P}_\Xi$ is the $\Xi$
polarization, $\alpha$ is the $\Xi\rt\Lambda\pi$ weak decay parameter and $\theta$ is the $\Lambda$
decay angle. When averaged over $\Theta_\Xi$, $\langle{\mathcal P}_{z^\prime}\rangle = \alpha$.
For both $\Xi^-$ and $\Xi^0$, the $\alpha$ values are substantial:
   for $\Xi^-\rt\Lambda\pi^-$ $\alpha_- =-0.458\pm 0.012$;
   for $\Xi^0\rt\Lambda\pi^0$ $\alpha_0 =-0.406\pm 0.013$.
In both cases, the average $\Lambda$ polarization is more than three times that for $\Lambda$s
from $\jpsi\rt\LLb$, as mentioned in the previous subsection.

The $A_{CP}$ sensitivity for $\Lambda$ decay is proportional to
$\langle{\mathcal P}\rangle\times \sqrt{N_{\rm evts}}$. The three times higher $\langle{\mathcal P}\rangle$
value means that a $\Lambda$ originating from $\jpsi\rt\Xi\bar{\Xi}$ is {\it nine times} more valuable than one
from $\jpsi\rt\LLb$ decays. Thus, even though the $\jpsi\rt\Xi\bar{\Xi}$ branching fractions are about
half that for $\jpsi\rt\LLb$, and the $\Xi\bar{\Xi}$ reconstruction efficiencies are also only about
half as large, the overall $A_{CP}$ sensitivity is almost three times better. This improved sensititivity
has been confirmed by detailed simulations~\cite{kupsc} and it is expected that BESIII's ultimate
$A_{CP}$ sensitivity for the existing 10B $\jpsi$ event sample will reach the $10^{-3}$ level.

\section{Three comments}
\noindent
{\bf 1.} Most of the physics program described above was not mentioned in the 850 page ``Physics at BESIII''
tome that we prepared prior to BESIII operation. These topics are just a few examples of the rich variety of
physics contained in the BESIII data.  After ten years of operation, we still discover
new and unexpected research opportunities with $\EE$ data in the 2-5~GeV energy range.

\noindent
{\bf 2.} After 100 years of research, there is still lots of physics to learn from the nucleons and
(after 50 years of research) hyperons. The time-like form-factor measurements discussed here are just
scratching the surface of what appears to be interesting and mostly unexplored physics.

\noindent
{\bf 3.} The search for matter-antimatter asymmetries with hyperons is the last frontier for SM $CP$ physics.
For this, exclusive $\EE\rt\jpsi\rt \Xi \bar{\Xi}$ processes are uniquely well suited.

\section{Personal remarks}
\noindent
While this symposium celebrates the 30$^{\rm th}$ anniversary of BES, it also marks the 35$^{\rm th}$ year
of my close connection to BES and IHEP. I first visited IHEP in 1984 to attend the first BES physics workshop.
During that visit I arranged with then Director-General Ming-Han Ye and researcher Zhi-Peng Zheng
the participation of Zhi-Peng's group in the AMY experiment at Tristan. In 1992, after eight fruitful
years of collaboration on AMY, Zhi-Peng invited me to join BES, which I did.
(I then recruited Fred Harris, which turned out to be my biggest contribution to BES.)
Our work together over these 35 years has been a long, often difficult, but always interesting trip.
In 1984, IHEP and BES were poor and struggling. Today IHEP is a world
premier laboratory for particle physics, and BESIII is a highly regarded, world-class experiment.
The whole story is much too complex to summarize succinctly in a report like this.
Maybe some day I'll write a book.

\begin{figure}[h]
\begin{center}
\includegraphics[width=5.0in]{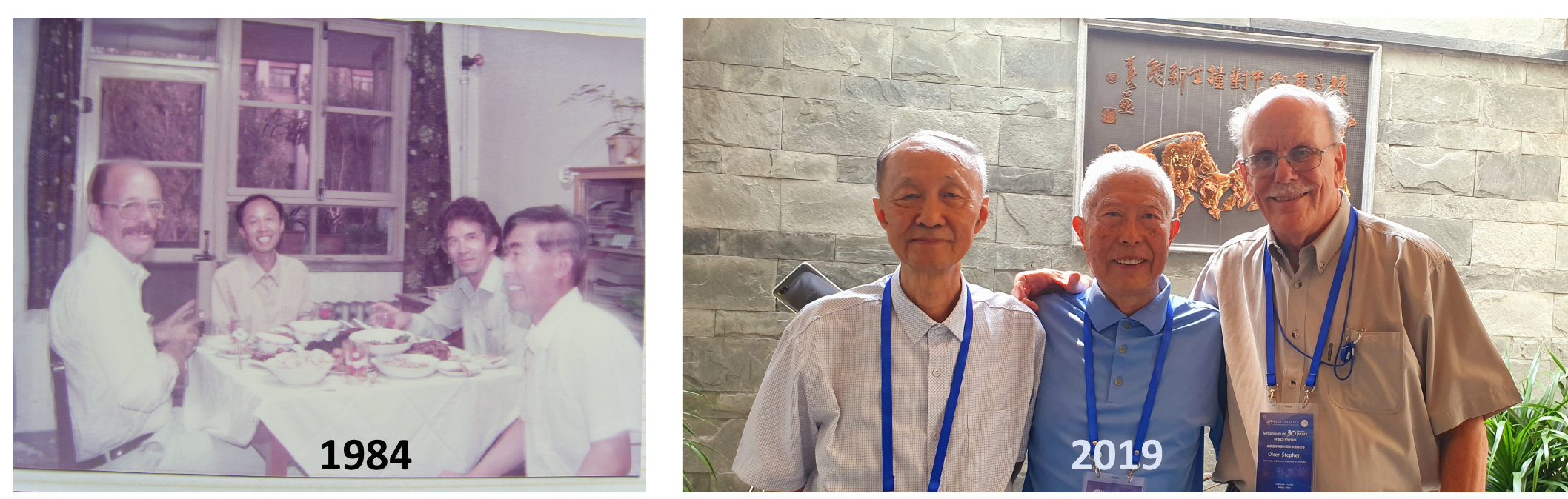}
\end{center}
\caption{ {\bf Left} Dinner with Zhi-Peng Zheng, an unidentified DESY physicist and Yu-Can Zhu at
  Zhi-Peng's apartment in 1984. {\bf Right} A 35$^{\rm th}$ year reunion at this symposium.}
\label{fig:photo}
\end{figure}

\end{document}